\documentclass[11pt]{article}

\usepackage[left=2.5cm,top=4cm,right=2.5cm,bottom=3cm]{geometry}
\usepackage{amsmath,amssymb}
\usepackage{placeins}
\usepackage{float}

\usepackage{subfigure,epsfig,epsf,color}
\usepackage[utf8x]{inputenc}
\usepackage{slashed}
\usepackage{epstopdf}
\usepackage{subfigure}
\usepackage{amsfonts}
\usepackage{cite}
\usepackage{pdflscape}
\usepackage{lipsum}
\begin{document}
\date{}

\title{Inflationary Cosmology with a scalar-curvature mixing term $\xi R \phi^2$ }
\maketitle
\begin{center}
Payel Sarkar\footnote{p20170444@goa.bits-pilani.ac.in},~Ashmita\footnote{p20190008@goa.bits-pilani.ac.in},~Prasanta Kumar Das\footnote{pdas@goa.bits-pilani.ac.in} \\
\end{center}
 
\begin{center}
Birla Institute of Technology and Science-Pilani, K. K. Birla Goa campus, NH-17B, Zuarinagar, Goa-403726, India
\end{center}
\vspace*{0.25in}
\begin{abstract}
\noindent We use the PLANCK 2018 and the WMAP data to constraint inflation models driven by a scalar field $\phi$ in the presence of the non-minimal scalar-curvature mixing term $\frac{1}{2}\xi R \phi^2$. We consider four distinct scalar field potentials $\phi^p e^{-\lambda\phi},~(1 - \phi^{p})e^{-\lambda\phi},~(1-\lambda\phi)^p$ and $\frac{\alpha\phi^2}{1+\alpha\phi^2}$ to study inflation in the non-minimal gravity theory. We calculate the potential slow-roll parameters and  predict the scalar spectral index $n_s$ and the tensor-to-scalar ratio $r$, in the parameters ($\lambda, p, \alpha$) space of the potentials. We have compared our results with the ones existing in the literature, and this indicates the present status of the non-minimal inflation after the release of the PLANCK 2018 data. 
\end{abstract}


\section{Introduction}
In the very beginning, the universe was dominated by dark energy with great density. It is the repulsive gravity of dark energy that made the universe expand with great exponential acceleration. This state of extremely dense dark energy lasted for about $10^{-33}$ s \cite{Liddle}. By this time, the universe has expanded so much that the distances between any two reference points must have increased by $50$-$60$ e-folds. This accelerating phase of the universe expansion is called the inflation\cite{KT, Guth, Linde1, Linde2, PS}. 
In addition to explaining why the universe is isotropic and homogeneous on a large scale, inflation elegantly resolves cosmological conundrums like the flatness problem and the horizon problem. A successful model of inflation requires a scalar field in a flat potential that rolls so slowly over a sufficiently long period during inflation \cite{KT, Guth,Amruth,Kurek,Weinberg,VSahni,Zlatev,Hrycyna,Linde2, Liddle, Baumann, Kinney}. It acts as the origin of density fluctuation which later evolves into the large-scale structure formation of the universe. The prediction of almost scale-invariant cosmological perturbation by inflation agrees remarkably well with the CMBR observations of COBE\cite{smoot}, WMAP\cite{WMAP}, PLANCK\cite{Planck} and BICEP2\cite{BICEP}. The WMAP data measures the spectral index of the scalar fluctuations $n_s = 0.99 \pm 0.04$ and put the $95\%$ CL upper limit on the tensor-to-scalar ratio, $r < 0.9$. The recent PLANCK-2018 mission measures the scalar spectral index $n_s = 0.9649 \pm 0.0042$ and put the  upper limit on $r < 0.1$ (at $95\%$ CL), which is further tightened by combining with the BICEP2/Keck Array BK15 data to obtain $r < 0.056$. 
\par
To study inflation, the commonly used potential is quadratic but the observational data no longer favor the basic inflationary model with a minimally coupled scalar field for a power law type potential as the tensor-to-scalar ratio predicted by a power law inflaton potential is quite high, which shows that this paradigm has to be extended. The simplest way to extend the scalar field Lagrangian is to include a non-minimal coupling term ($\xi$) between the curvature ($R$) and the scalar field of the form $\frac{1}{2} \xi R\phi^2$ \cite{Accetta, Girish, Hertzberg, Orest, Faraoni, Yuttana, Sergey, Burin, Muta, Odintsov, Mario, Yi, Kallosh, Yi2, capovilla, aatifa, Nojiri, Inagaki, fakir, Futamase}. The coupling constant $\xi$ in this scenario becomes a free parameter in the model from the theoretical point of view and an effective theory approach.
Its value should be estimated from the observational data by taking a pragmatic approach.\\
\noindent
Two values of $\xi$ are often discussed in the literature, one having $\xi=0$ i.e. minimal case while the other is conformal coupling where $\xi=1/6$ \cite{Birrell, Sonego}. Contrary to popular assumption, there are numerous strong arguments for including non-minimal coupling, which is of relevance from a physical and cosmological perspective\cite{Gunzig}.
It arises at the quantum level when quantum corrections to the scalar field theory are considered. It is required for the renormalization\cite{Buchbinder, Elizalde} of the scalar field theory in curved space\cite{Allen, Ishikawa, Davis, Parker, Callan}. The problem of choosing the value of $\xi$ has been addressed at both classical and quantum levels, it depends on the theory of gravity and the scalar field adopted. 
\noindent
The non-minimal inflation model describes the cosmic expansion with a graceful exit \cite{Nozari} towards its end\cite{cheong, Jin}.
It is also intriguing to consider non-minimal coupling scenarios in the context of multidimensional theories like super-string theory \cite{Maeda} and induced gravity\cite{Accetta2}.\\
In this work, we analyze a non-minimal inflation model with a scalar-curvature mixing term\cite{Gron} $ \frac{1}{2}\xi R\phi^2$ using four different types of scalar potentials (i) $V = V_0\phi^pe^{-\lambda\phi}$, (ii) $V = V_0(1-\phi^{p})e^{-\lambda\phi}$, (iii) $V=V_0(1-\lambda\phi)^p$ and (iv) $V=V_0\frac{\alpha\phi^2}{1+\alpha\phi^2}$, where $\lambda, p$ and $\alpha$ are the potential parameters.
In this regard, a significant amount of work has been done in literature for several potentials. However, the first and second potentials are entirely new in the context of both minimal and non-minimal inflation (to the extent we know).
The third and fourth potentials have been investigated by Gr{\o}n\cite{Gron} in the minimally coupled gravity theory and no one has studied these potentials in the non-minimal inflationary theory. 
The first two potentials are the product of power law and exponential term, although the power law and the exponential potential have been thoroughly explored in literature\cite{Nozari, Tsujikawa, Zarei} independently while the combination is not and in both cases the tensor-to-scalar ratio is quite high ($r \approx 0.256$\cite{Gron}) in the minimally coupled gravity. In order to get the spectral parameters of CMB to match the observational data, we are using a mix of a power law and an exponential term in non-minimally coupled gravity. It turns out that the coupling constant must have a non-vanishing value to depict the inflationary paradigm. We have also studied the slow-roll inflation in the context of modified gravity for a similar set of potentials in \cite{Ashmita}.
\par  
The paper is organized as follows: In Section 2, we derive the Friedmann equations and the scalar field equations with the scalar field coupled non-minimally to gravity. In Section 3, we obtain the slow-roll parameters for four distinct potentials in the non-minimal gravity theory, derive the spectral index parameters $n_s$, $r$, and obtain constraints on them in the parameter space of the potential for different $\xi$ values using the WMAP and PLANCK 2018 data. Then we discuss the cosmological viability of our model. Finally, in Section 4, we summarize our results and conclude.

\section{Non-minimal Inflation in Einstein Frame:}

The action for gravity with a non-minimally coupled scalar field in the Jordan frame is given by
\begin{equation}
\label{action}
     S_J = \int d^4 x ~ \sqrt{-g} \left[\frac{1}{2 \kappa^2} R - \frac{1}{2} g^{\mu\nu}~\partial_\mu{\phi}~\partial_\nu{\phi} + V(\phi) +\frac{1}{2} \xi R \phi^2  \right]
\end{equation}
Here $R$ is the Ricci scalar, $\kappa^2 = 8 \pi G_N = 8 \pi/M^2_{pl}$\footnote{Here, we set $c = 1, 8\pi G=1$} and $g$ is the determinant of the metric $g_{\mu\nu}(x)$.
$V(\phi)$ is the potential of the scalar field and $\xi$ is the non-minimal coupling of the scalar field $\phi$ with $R$. The metric sign convention is chosen to be $(+,-,-,-)$ with spatially flat Friedman-Robertson-walker metric as,

\begin{equation}
    ds^2=dt^2-a^2(t)(dx^2+dy^2+dz^2)
\end{equation}
To get the FRW equations in the Einstein frame, we perform a conformal transformation (Weyl's rescaling)  \cite{Kaiser, Nozari, Qiu, Postma},

\begin{equation}
    \hat{g}_{\mu\nu}(x)=\Omega^2(x)g_{\mu\nu}(x)
\end{equation}
where the hat on variables is used in Einstein's frame. In the Einstein frame, the Christoffel symbol will transform as \cite{Kaiser,dabrowski}

\begin{equation}
    \hat{\Gamma}^{\alpha}_{\beta\gamma}=\Gamma^{\alpha}_{\beta\gamma}+\frac{1}{\Omega}\left(\delta^{\alpha}_{\gamma}\partial_{\beta}\Omega+\delta^{\alpha}_{\beta}\partial_{\gamma}\Omega-\partial^{\alpha}\Omega g_{\beta\gamma}\right)
\end{equation}
Similarly, the transformation of the Ricci tensor and Ricci scalar will be,

\begin{equation}
    \hat{R}_{\mu\nu}=R_{\mu\nu}+\frac{1}{\Omega^2}\left(4\Omega_{,\mu}\Omega_{,\nu}-\Omega_{,\alpha}\Omega^{,\alpha}g_{\mu\nu}\right)-\frac{1}{\Omega}\left(2\Omega_{;\mu\nu}+g_{\mu\nu}\Box\Omega\right),
  \end{equation}

  \begin{equation}
     \hat{R}=\frac{1}{\Omega^2}\left (R-\frac{6\Box\Omega}{\Omega}\right)
\end{equation}
Here, $\sqrt{-\hat{g}}=\Omega^4\sqrt{-g}$. Considering, $\Omega^2(x)=1+\kappa^2 \xi\phi^2$ (with $\kappa^2 = 1$), we can write the action in Einstein frame as 

\begin{equation}
    S_E=\int d^4x\sqrt{-\hat{g}}\left(\frac{1}{2\kappa^2}\hat{R}-F^2(\phi)\frac{1}{2}\hat{g}^{\mu\nu}\hat{\partial}_{\mu}\phi\hat{\partial}_{\nu}\phi+\hat{V}(\phi)\right)
\end{equation}
where $\frac{d\hat{\phi}}{d\phi}=F(\phi)=\frac{\sqrt{1+\xi\phi^2(1+12\xi)}}{1+\xi\phi^2}$ and $\hat{V}(\phi)= \Omega^{-4}~ V(\phi)$. Note that in the Einstein frame, the scalar field $\phi$ is no longer coupled with the Ricci scalar $R$. The invariance of the $4$-dimensional line element under the Conformal Transformations gives

 \begin{equation}
   ds^2 = g_{\mu\nu} dx^\mu dx^\nu = \Omega^{-2} \hat{g}_{\mu\nu} d{\hat x}^\mu d{\hat x}^\nu =  \Omega^{-2} d\hat{s}^2 \to  d\hat{s}^2=d\hat{t}^2-\hat{a}^2(\hat{t})(d\hat{x}^2+d\hat{y}^2+d\hat{z}^2)
 \end{equation}
where $\hat{a}(t)=\Omega(x)~ a(t)$. \\
The energy-momentum tensor in Einstein's frame is found to be 
\begin{eqnarray} \label{tmunu}
   \hat{T}_{\mu\nu}=\hat{g}_{\mu\nu}\left(-F^2(\phi)\frac{1}{2}\hat{\partial}_{\alpha}\phi\hat{\partial}^{\alpha}\phi+\hat{V}\right)+F^2(\phi)\hat{\partial}_{\mu}\phi\hat{\partial}_{\nu}\phi
\end{eqnarray}
The Friedmann equation and the scalar field equation in Einstein frame\cite{Nozari} can be written as

\begin{equation}
\label{hubble}
    \hat{H}^2=\frac{1}{3} \hat{\rho}_\phi = \frac{1}{6}\left(\frac{d\hat{\phi}}{d\hat{t}}\right)^2+\frac{1}{3}\hat{V}(\hat{\phi})
\end{equation}

\begin{equation}
\label{scalarEoS}
~ ~ \frac{d^2\hat{\phi}}{d\hat{t}^2}+3\hat{H}\frac{d\hat{\phi}}{d\hat{t}}+\frac{d\hat{V}}{d\hat{\phi}}=0
\end{equation}
\noindent
Under the slow-roll approximations $\dot{\hat{\phi}}^2<<\hat{V}(\hat{\phi}), \ddot{\hat{\phi}}<<3\hat{H}\dot{\hat{\phi}}$ (where $\dot{\hat{\phi}} = \frac{d\hat{\phi}}{d\hat{t}}$ and $ \ddot{\hat{\phi}} = \frac{d^2 \hat{\phi}}{d{\hat t}^2}$), the Hubble equation (\ref{hubble}) and the scalar field equations(\ref{scalarEoS}) take the following form 

\begin{equation}
    \hat{H}^2=\frac{1}{3}\hat{V}(\hat{\phi}),
\end{equation}
\begin{equation}
    3\hat{H}\frac{d\hat{\phi}}{d\hat{t}}+\frac{d\hat{V}}{d\hat{\phi}}=0
\end{equation}

\section{Analysis of non-minimal inflation with a class of scalar potentials}

We next derive the potential slow-roll parameters ($\epsilon_v,~\eta_v$), e-fold number ($N$), and CMBR observables like scalar spectral index and tensor-to-scalar ratio in the case of four scalar potentials and obtain constraints on them in the parameter space of the potentials using the observational data.

\subsection{Slow roll parameters and CMB constraints:}
We presumptively consider that the universe is filled with a spatially homogenous scalar field and the universe must have passed through an inflationary phase of expansion. In order to get this phase, the potential energy term of the scalar field Lagrangian must prevail over the kinetic energy term i.e. $V(\phi) >>\frac{1}{2}\dot{\phi}^2 $. This is also known as the {\it slow-roll condition}. To study inflation, we start by defining the potential slow roll parameters\cite{Nozari} in the Einstein frame as follows
\begin{eqnarray}\label{slow roll}
\epsilon_v = \frac{1}{2}\left(\frac{\hat{V}'(\phi)}{\hat{V}(\phi)}\right)^2,  
~~\eta_v = \frac{\hat{V}^{''}(\phi)}{\hat{V}(\phi)}
\end{eqnarray}
where, 
\begin{equation*}
    \hat{V}^{'}(\phi) = \frac{d\hat{V}}{d\hat{\phi}} = \frac{d\hat{V}}{d\phi} \frac{d\phi}{d\hat{\phi}} = \frac{d\hat{V}}{d\phi} \frac{1}{F(\phi)}, ~~ \hat{V}^{''}(\phi)=\frac{d\hat{V}^{'}}{d\phi}\frac{d\phi}{d\hat{\phi}} = \frac{1}{F(\phi)} \frac{d}{d\phi} \left(\frac{d\hat{V}}{d\phi} \frac{1}{F(\phi)}\right)
\end{equation*}
The slow-roll parameters are related to the CMB observables - the scalar spectral index, tensor spectral index and tensor-to-scalar ratio as follows,
\begin{equation}\label{spectral index}
    n_s-1=2\eta_v-6\epsilon_v, ~~ r=16 \epsilon_v, ~~n_T=-2\epsilon_v
\end{equation}
The e-fold number is defined as the ratio of the final value of the scale factor $a_f$ (at the time of exit from the inflation) and its initial value $a_i$ can be calculated as,

\begin{equation}
\label{e-fold}
 N= ln\left(\frac{a_f}{a_i}\right)= - \int_{\phi_{in}}^{\phi_f} \frac{\hat{V}}{{\hat V}_{,\hat \phi}} d {\hat \phi} =  -\frac{1}{\sqrt{2}}\int_{\phi_{in}}^{\phi_{f}}(\epsilon_v)^{-1/2}\frac{d\hat{\phi}}{d\phi}d\phi
\end{equation}

\subsubsection{Case 1: $V=V_0 \phi^p e^{-\lambda\phi}$}

We start with the potential (introduced first by {\it{Ashmita} et. al.})  \cite{Ashmita}), 
\begin{equation}
    V = V_0 \phi^p e^{-\lambda\phi}
    \label{potential}
\end{equation}
where $V_0$ is a constant, $p$ (the power index) and $\lambda$ are the potential parameters. We find that for $\lambda=0$, this potential reduces to ``chaotic potential $(\phi^p)$". On the other hand, $p=0$ leads to  exponential inflationary potential with positive curvature discussed in \cite{Valentino}.
\\
Now, in Einstein's frame, the potential can be written as,
\begin{equation}
    \hat{V}(\phi)=\frac{V_0\phi^p e^{-\lambda\phi}}{\bigl(1+\xi\phi^2\bigr)^2}
\end{equation}
The potential slow-roll parameters can be obtained as 

\begin{equation}\label{epsilon1}
    \epsilon_v = \phantom{-} \frac{\Bigl[p+p\xi\phi^2-\phi\bigl(\lambda+4\xi\phi+\lambda
    \xi \phi^2\bigr)\Bigr]^2}{2\phi^2\bigl(1+\xi\phi^2+12\xi^2\phi^2\bigr)}
\end{equation}

\begin{equation}\label{eta1}
\begin{split}
     \eta_v
&=\frac{1}{\phi^2\bigl(1+\xi\phi^2+12\xi^2\phi^2\bigr)^2} \times \biggl[\bigl(p+p\xi\phi^2\bigr)^2\bigl(1+\xi\phi^2+12\xi^2\phi^2\bigr)-p\bigl(1+\xi\phi^2\bigr)\Bigl\{1+ \\
&\quad 9\xi\phi^2+96\xi^3\phi^4+8\xi^2\phi^2\bigl(3+\phi^2\bigr)+2\lambda\bigl\{\phi+2\xi\phi^3\bigl(1+6\xi\bigr)+\xi^2\phi^5\bigl(1+12\xi\bigr)\bigr\}\Bigr\}+\\
&\quad \phi^2\bigl(\lambda+\lambda\xi\phi^2\bigr)^2\bigl(1+\xi\phi^2+12\xi^2\phi^2\bigr)+4\xi\bigl(-1+3\xi\phi^2+4\xi^2\phi^4+48\xi^3\phi^4\bigr)+\lambda\xi\phi \times \\
&\quad \Bigl\{7+84\xi^3\phi^4+2\xi\bigl(6+7\phi^2\bigr)+\xi^2\phi^2\bigl(96+7\phi^2\bigr)\Bigr\}\biggr]
\end{split}
\end{equation}
Note that $n_s$, $r$, and $n_T$, the three spectral index parameters, are independent of the frame of reference i.e. conformally invariant \cite{Makino}.
From Eq.~(\ref{epsilon1}), Eq.~(\ref{eta1}) and Eq.~(\ref{spectral index}) we evaluate $n_s$ as 

\begin{equation}\label{ns1}
    \begin{split}
        n_s
        &=-\frac{1}{\phi^2\bigl(1+\xi\phi^2+12\xi^2\phi^2\bigr)^2}\biggl[\bigl(p+p\xi\phi^2\bigr)^2\bigl(1+\xi\phi^2+12\xi^2\phi^2\bigr)-2p\bigl(1+\xi\phi^2\bigr)\Bigl\{-1\\
        &\quad +3\xi\phi^2+48\xi^3\phi^4+4\xi^2\phi^2\bigl(\phi^2-6\bigr)+\lambda\bigl\{\phi+2\xi\phi^3\bigl(1+6\xi\bigr)+\xi^2\phi^5\bigl(1+12\xi\bigr)\bigr\}\Bigr\}+\\
       &\quad
       \phi^2\Bigl\{-1-\xi^2\phi^4-8\xi^3\phi^4+48\xi^4\phi^4-2\xi\bigl(-4+\phi^2\bigr)+\bigl(\lambda+\lambda\xi\phi^2\bigr)^2\bigl(1+\xi\phi^2+\\
       &\quad 
       12\xi^2\phi^2\bigr)+2\lambda\xi\phi\bigl\{5+60\xi^3\phi^4+2\xi\bigl(-6+5\phi^2\bigr)+\xi^2\phi^2\bigl(48+5\phi^2\bigr)\bigr\}\Bigr\}\biggr]
    \end{split}
\end{equation}
and $r$, $n_T$ as follows,
\begin{equation}
    r=\frac{8\Bigl[p+p\xi\phi^2-\phi\bigl(\lambda+4\xi\phi+\lambda\xi\phi^2\bigr)\Bigr]^2}{\phi^2\bigl(1+\xi\phi^2+12\xi^2\phi^2\bigr)}, ~~     n_T=-\frac{\Bigl[p+p\xi\phi^2-\phi\bigl(\lambda+4\xi\phi+\lambda\xi\phi^2\bigr)\Bigr]^2}{\phi^2\bigl(1+\xi\phi^2+12\xi^2\phi^2\bigr)}
\end{equation}
\noindent
In Table (\ref{table:1}), we have tabulated the values of  $n_s$ and $r$ for e-fold number $N=50$ and $60$, respectively for two different values of non-minimal parameter $\xi(=0.001, 0.0001)$. A successful inflationary model should typically generate the required e-fold before inflation ends in order to solve the horizon and flatness problems.
\begin{table}[H]
\addtolength{\tabcolsep}{9.2pt}
 \centering
 \small
 \begin{tabular}{ccccccccc}
 \hline
  $\lambda$ & p & $\xi$ & $\phi_f$ & $N$ & $\phi$ & $n_s$ & $r$\\
  \hline
  0 & 2 & 0.001 & 1.40998 & 50 & 13.519 & 0.959039 & 0.098686\\
    & & & & 60 & 14.66 & 0.965409 & 0.07537\\
    & 2 & 0.0001 & 1.41379 & 50 & 14.146 &  0.960407 & 0.15056\\
    & & & & 60 & 15.48 & 0.966997 & 0.12424\\
    \hline
    0.1 & 2 & 0.001 & 1.31729 & 50 & 10.352 & 0.949976 & 0.02755\\
    & & & & 60 & 10.906 & 0.954881 & 0.01763\\
    & 2 & 0.0001 & 1.32046 & 50 & 10.986 & 0.95974 & 0.04889\\
    & & & & 60 & 11.71 & 0.965367 & 0.03551\\
    \hline
    0.15 & 2 & 0.001 & 1.27537 & 50 & 9.035 & 0.939382 & 0.01245\\
    & & & & 60 & 9.388 & 0.94361 & 0.007077\\
    & 2 & 0.0001 & 1.27827 & 50 & 9.65 & 0.953083 & 0.02305\\
    & & & & 60 & 10.10 & 0.957736 & 0.01566\\
    \hline
    0.1 & 4 & 0.001 & 2.63183 & 50 & 16.376 & 0.949808 & 0.08678\\
    & & & & 60 & 17.459 & 0.956774 & 0.05947\\
    & 4 & 0.0001 & 2.64065 & 50 & 16.89 & 0.953391 & 0.139605\\
    & & & & 60 & 18.14 & 0.96125 & 0.106412\\
    \hline
    0.15 & 4 & 0.001 & 2.54805 & 50 & 14.768 & 0.946037 & 0.050915\\
    & & & & 60 & 15.56 & 0.951914 & 0.03217\\
    & 4 & 0.0001 & 2.55628 & 50 & 15.369 & 0.953763 & 0.08901\\
    & & & & 60 & 16.358 & 0.960854 & 0.063834\\
    \hline
 \end{tabular}
 \caption{\label{table:1} The e-fold number $N$, the spectral index parameters $n_s$ and $r$ are presented here for the potential $V=V_0\phi^{p}e^{-\lambda\phi}$.}
 \end{table}

 We have calculated the field value at the end of inflation $\phi_f$ by considering $\epsilon_v$ to be 1 and obtained the field value $\phi$ (initial seed value of the field necessary to trigger the inflation) by taking $N=50$ and $60$ as input.
We can observe that, for chaotic potential i.e. $(p=2,\lambda=0)$, $r$ is quite high for $\xi=0.0001$.
In contrast, by including the exponential term i.e. for non-zero $\lambda$ values, we are getting a desirable range of $n_s$ and $r$, which provides the justification for adopting a mix of exponential and power-law potential.
\begin{figure}[H]
 \centerline{\psfig{file=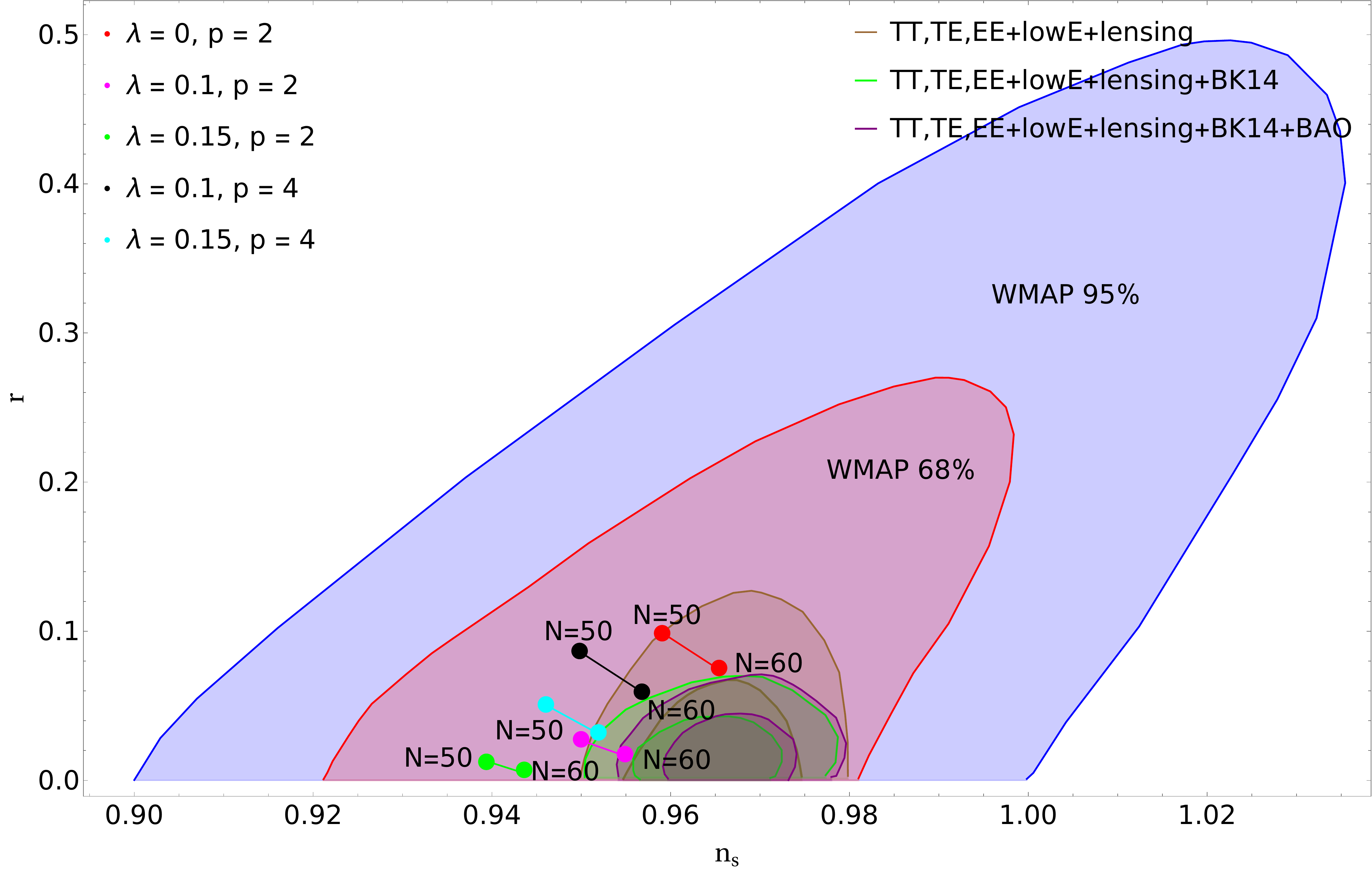, width=8.5cm}\psfig{file=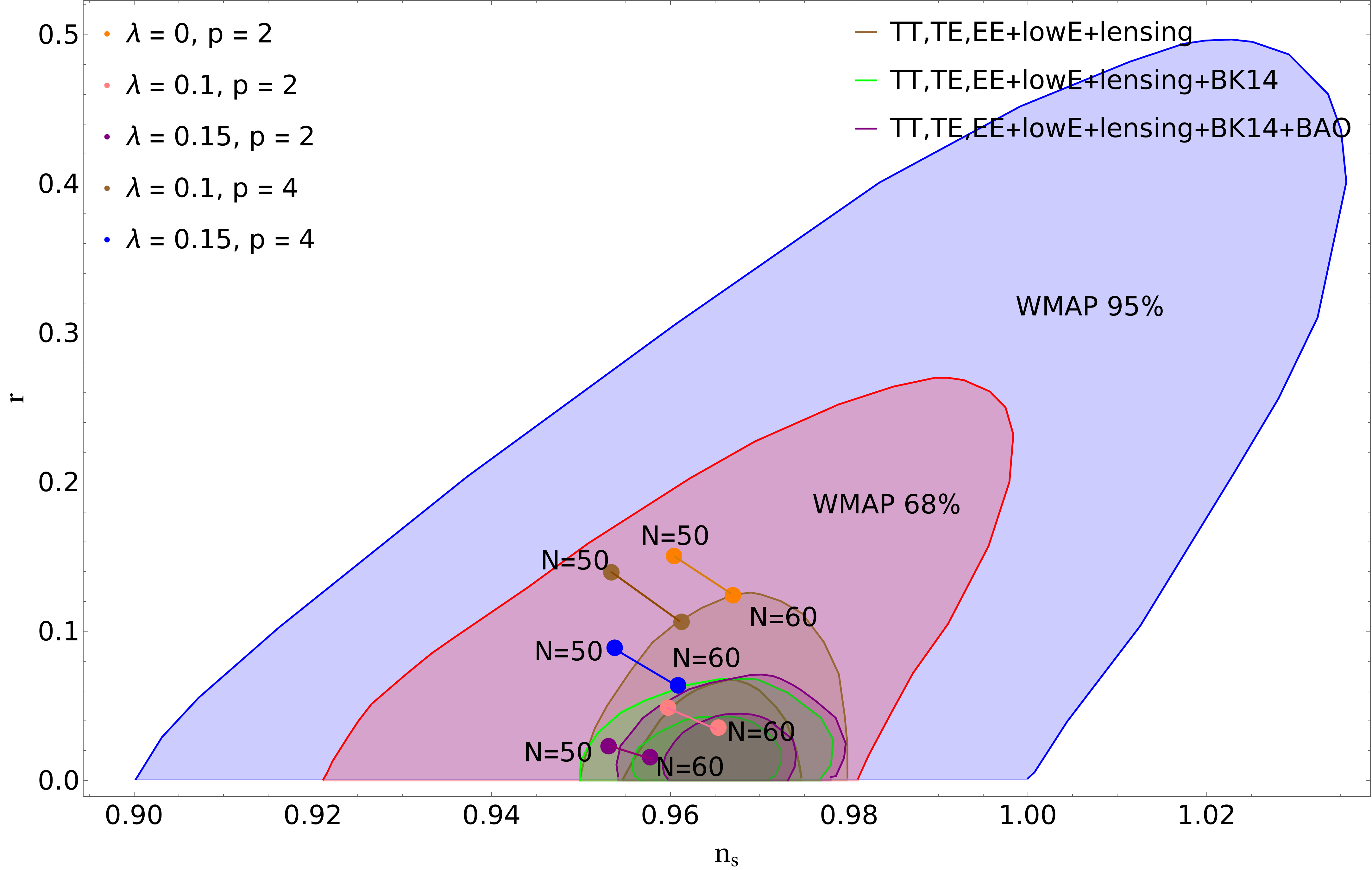, width=8.5cm}}
  \vspace*{5pt}
  \caption{\label{figure1}{(Color online) Constraints on $n_s$ and $r$ from CMB measurements for the potential $V=V_0\phi^{p}e^{-\lambda\phi}$ for different values of p and $\lambda$. Shaded regions are allowed by WMAP measurements, PLANCK alone, PLANCK+BK15, PLANCK+BK15+BAO to $68\%$ and $95\%$ confidence level.}} 
  
\end{figure}
 \noindent
For the aforementioned potential, we have shown the variation of $n_s$ and $r$ in Fig (\ref{figure1}). The blue and red shaded regions correspond to WMAP data upto $95\%$ and $68\%$ C.L. respectively whereas brown, green, and purple shaded regions correspond to PLANCK, PLANCK+BK15, PLANCK+BK15+BAO where the two same color curves in the figure denote the $68\%$ and $95\%$ region for Planck data. 
On the left-hand side of Fig (\ref{figure1}), the e-fold number $N=50$ and $60$ corresponds to $\xi=0.001$ for potential $V=V_0\phi^pe^{-\lambda\phi}$ with (i) $p=2$, $\lambda=0$, $0.1$, $0.15$ and (ii) $p=4$, $\lambda=0.1$, $0.15$ respectively are shown in red, pink, green, black, cyan whereas the right side of the plot is shown for $\xi=0.0001$ with the same set of parameter choices of $p$ and $\lambda$ shown in orange, pink, purple, brown and blue colors. 
For $\xi=0.001$, only $\lambda=0$, $p=2$ and $\lambda=0.1$, $p=2$ lies within the $\pm 3 \sigma$ range of the scalar spectral index data, and the tensor-to-scalar ratio is also allowed by the Planck 2018 data. 
With the increase in $p$-value, we see that $n_s $ goes beyond the $3\sigma$ limit and is ruled out according to the observational data. 
Even for $\lambda=0.15$ and $p=2$, we still get $n_s$ value greater and is ruled out according to the Planck data. On the other hand, by lowering the value of $\xi$ to $0.0001$, we see that $\lambda=0.1$, $p=2$ and $\lambda =0.15$, $p=2$ are the best choices that fit with the Planck data while the other choices are discarded. As for other cases, either it does not follow the $\pm 3 \sigma$ bound limit on $n_s$ or the tensor-to-scalar ratio is quite high. \\   
\subsubsection{Case 2: $V=V_0(1-\phi^{p})e^{-\lambda\phi}$}

\begin{figure}[H]
 \centerline{\psfig{file=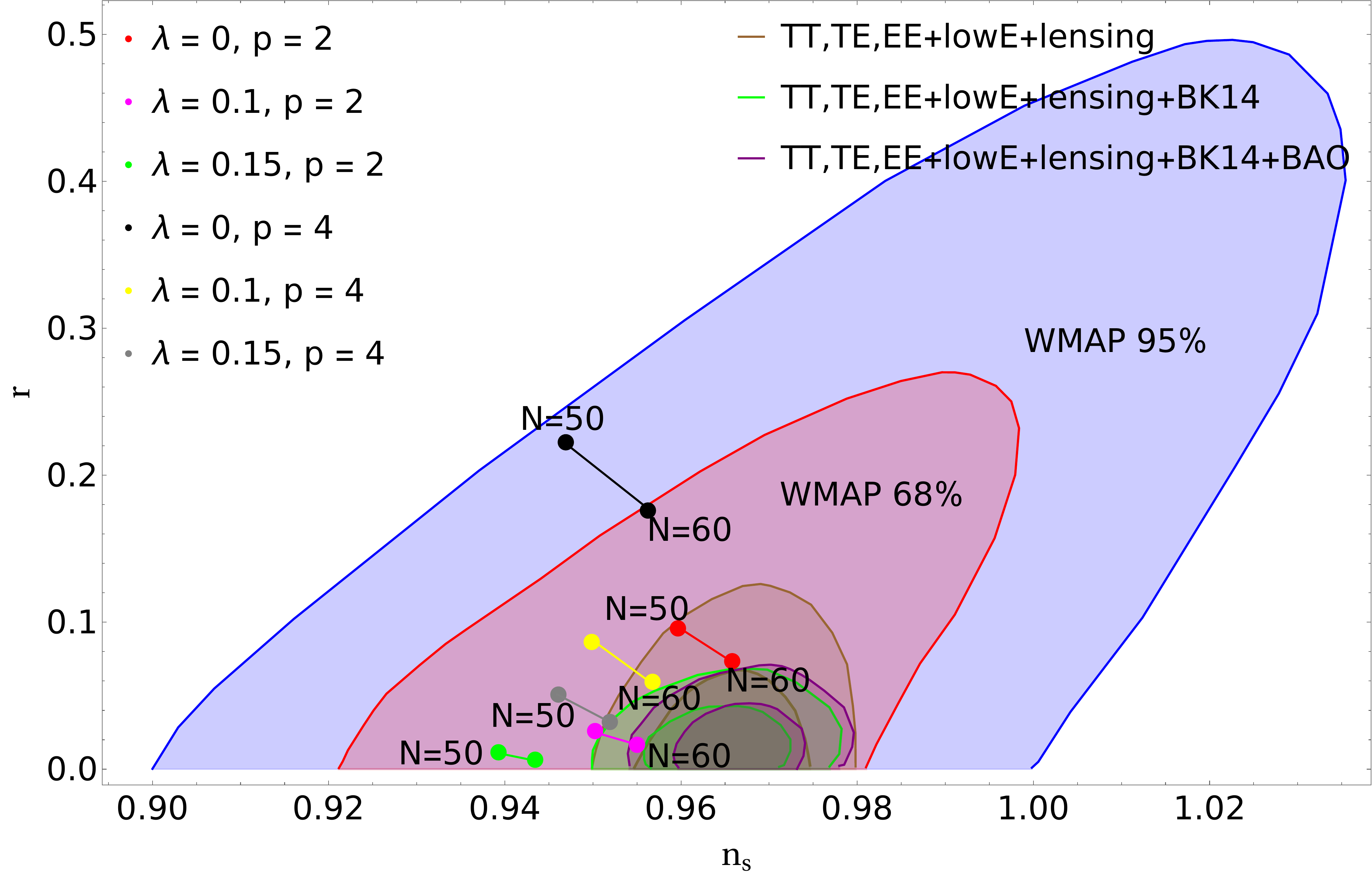, width=8.5cm}\psfig{file=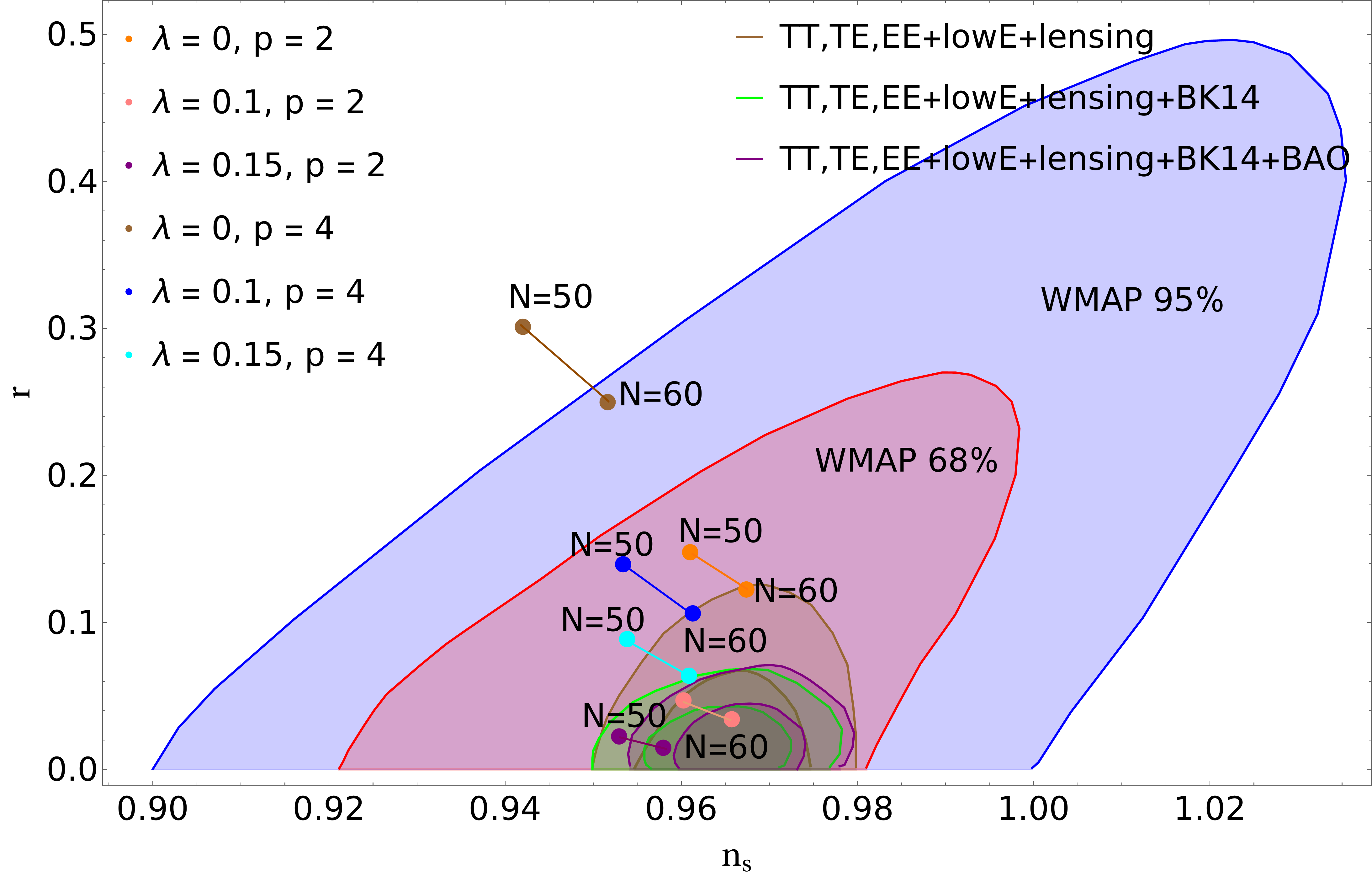, width=8.5cm}}
  \vspace*{5pt}
  \caption{\label{figure2}{(Color online) Constraints on $n_s$ and $r$ from CMB measurements for the hilltop potential along-with an exponential term $V=V_0(1-\phi^{p})e^{-\lambda\phi}$ for $\xi=0.0001$ (left side) and $\xi=0.001$ (right side). Shaded regions are allowed by WMAP measurements, PLANCK alone, PLANCK+BK15, PLANCK+BK15+BAO to $68\%$ and $95\%$ confidence levels. 
  }}
\end{figure}

Next, let us turn on to the potential $ V(\phi) = V_0 \bigl(1 - \phi^p\bigr)e^{-\lambda \phi}$. In the Einstein frame, it takes the form
\begin{equation}
\hat{V}(\phi) = \frac{V_0 \bigl(1 - \phi^p\bigr)e^{-\lambda \phi}}{\bigl(1 + \xi \phi^2\bigr)^2}
\end{equation}
For $\lambda=0$ and $p=2$ the potential reduces to ``Hilltop potential" and the Inflationary model with this potential is known as ``{\it New Inflation}" which is discussed in \cite{Lyth} for the minimal scenario. Here, we have considered a more generalized version of hilltop potential by including an exponential factor in the non-minimal inflation. \\
\noindent We derive the potential slow-roll parameters for this  potential as follows
\begin{equation}\label{ns2}
    \epsilon_v =\frac{\Bigl[p \phi^p -\lambda \phi \bigl(1+\xi \phi^2\bigr)\bigl(-1+\phi^p\bigr) + \xi \phi^2 \bigl(4+(-4+p)\phi^p\bigr)\Bigr]^2}{2\phi^2 \bigl(1+\xi \phi^2+12 \xi^2 \phi^2\bigr) \bigl(-1+\phi^p\bigr)^2}
\end{equation}
\begin{equation}\label{eta2}
\begin{split}
    \eta_v 
    &=-\frac{1}{\bigl\{\phi+\xi(1+12\xi)\phi^3\bigr\}^2\bigl(1-\phi^p\bigr)} \times \biggl[\bigl(-1+p\bigr)p\phi^p + \lambda^2 \phi^2\bigl(1+\xi \phi^2\bigr)^2\bigl(1+\xi \phi^2+ \\
    &\qquad 12\xi^2\phi^2\bigr)\bigl(-1+\phi^p\bigr)+12\xi^4\phi^6\bigl(-16+(-4+p)^2\phi^p\bigr)+\xi \phi^2\bigl\{4+\bigl(-4-10p+3\\
    &\qquad \times p^2\bigr) \phi^p\bigr\}+\xi^3\phi^4\bigl\{-16 \phi^2+24\bigl(-5+p\bigr)p\phi^p+\bigl(-4+p\bigr)^2\phi^{2+p}\bigr\}-\xi^2\bigl\{12\phi^4- \\
    &\qquad\ 12\bigl(-2+p\bigr)p\phi^{2+p}+\bigl(-12+17p- 3p^2\bigr)\phi^{4+p}\bigr\}-\lambda\phi\bigl(1+\xi \phi^2\bigr)\Bigl\{2p\phi^p +12\xi^3\phi^4\\
    &\qquad \times \bigl\{7+\bigl(-7+2p\bigr)\phi^p\bigr\}+\xi\phi^2\bigl\{7+\bigl(-7+4p\bigr)\phi^p\bigr\}+\xi^2\phi^2 \bigl\{12+7\phi^2+12(-1+\\
    &\qquad 2p)\phi^p+\bigl(-7+2p\bigr)\phi^{2+p}\bigr\}\Bigr\}\biggr]
\end{split}
\end{equation} 
 Using Eq. (\ref{spectral index}), Eq. (\ref{ns2}) and Eq. (\ref{eta2}), we obtain the scalar spectral index as 
\begin{equation}
\begin{split}
     n_s 
     &=1-\frac{3\Bigl\{p \phi^p -\lambda \phi\bigl(1+\xi \phi^2\bigr)\bigl(-1+\phi^p\bigr)+\xi \phi^2\bigl\{4+\bigl(-4+p\bigr)\phi^p\bigr\}\Bigr\}^2}{\phi^2 \bigl(1+\xi \phi^2 +12 \xi^2 \phi^2\bigr) \bigl(-1+\phi^p\bigr)^2}-\\
     &\qquad \frac{2}{\bigl\{\phi +\xi\bigl(1+12\xi\bigr)\phi^3\bigr\}^2\bigl(1-\phi^p\bigr)} \biggl[\bigl(-1+p\bigr)p\phi^p+\lambda^2\phi^2\bigl(1+\xi\phi^2\bigr)^2\bigl(1+\xi\phi^2+12 \times \\
     &\qquad \xi^2\phi^2\bigr) \bigl(-1+\phi^p\bigr)+12\xi^4\phi^6\bigl\{-16+\bigl(-4+p\bigr)^2\phi^p\bigr\}+\xi\phi^2\bigl\{4+\bigl(-4-10p+3p^2\bigr) \times \\
     &\qquad \phi^p\bigr\}+\xi^3\phi^4\bigl\{-16\phi^2+24\bigl(-5+p\bigr)p\phi^p + \bigl(-4+p\bigr)^2\phi^{2+p}\bigr\}
     -\xi^2\bigl\{12\phi^4-12\bigl(-2+ \\
     &\qquad p\bigr)p\phi^{2+p}+ \bigl(-12+17p-3p^2\bigr)\phi^{4+p}\bigr\}-\lambda \phi\bigl(1+\xi \phi^2\bigr)\Bigl\{2p\phi^p+12\xi^3\phi^4\bigl\{7+\bigl(-7+\\
     & \qquad 2p\bigr)\phi^p\bigr\} + \xi \phi^2\bigl\{7+\bigl(-7+4p\bigr)\phi^p\bigr\}+\xi^2\phi^2\bigl\{12+7\phi^2+12\bigl(-1+2p\bigr) \phi^p+\bigl(-7+\\
     & \qquad 2p\bigr)\phi^{2+p}\bigr\}\Bigr\}\biggr]
\end{split}
\end{equation}
and the tensor-to-scalar ratio and the tensor spectral index as 
\begin{equation}
    r= \frac{8\Bigl[p \phi^p -\lambda \phi \bigl(1+\xi \phi^2\bigr)\bigl(-1+\phi^p\bigr) + \xi \phi^2 \bigl\{4+(-4+p)\phi^p\bigr\}\Bigr]^2}{\phi^2 \bigl(1+\xi \phi^2+12 \xi^2 \phi^2\bigr)\bigl(-1+\phi^p\bigr)^2}
\end{equation}
\begin{equation}
    n_T= -\frac{\Bigl[p \phi^p -\lambda \phi \bigl(1+\xi \phi^2\bigr)\bigl(-1+\phi^p\bigr) + \xi \phi^2 \bigl\{4+(-4+p)\phi^p\bigr\}\Bigr]^2}{\phi^2 \bigl(1+\xi \phi^2+12 \xi^2 \phi^2\bigr)\bigl(-1+\phi^p\bigr)^2}
\end{equation}
\noindent In Table (\ref{table:2}), we have calculated the values of $n_s$ and $r$ for a fixed value of $\phi$ which is obtained by taking $N=50$ and $60$. In Fig~(\ref{figure2}), we have shown the contour plot in the $n_s-r$ plane for the potential $V=V_0(1-\phi^p)e^{-\lambda\phi}$.
For $\xi=0.0001$, we can see the effect of using the combination of hilltop and exponential terms in the second potential. The best choice of parameters, in this case, are $\lambda=0.1$, $p=2$ and $\lambda=0.15$, $p=2$ for which both $n_s$ and r match well with the observational data given by Planck 2018. 
\begin{table}[H]
\addtolength{\tabcolsep}{7.6pt}
 \centering
 \small
 \begin{tabular}{ccccccccc}
 \hline
  $\lambda$ & p & $\xi$ & $\phi_f$ & $N$ & $\phi$ & $n_s$ & $r$\\
      \hline
    0 & 2 & 0.001 & 1.92785 & 50 & 13.71 & 0.95965 & 0.09580\\
    & & & & 60 & 14.83 & 0.96581 & 0.07346\\
    & 2 & 0.0001 & 1.93145 & 50 & 14.34 & 0.96100 & 0.14776\\
    & & & & 60 & 15.65 & 0.96738 & 0.12243\\
      \hline
    0.1 & 2 & 0.001 & 1.85543 & 50 & 10.515 & 0.950246 & 0.02594\\
    & & & & 60 & 11.053 & 0.95501 & 0.01661\\
    & 2 & 0.0001 & 1.85846 & 50 & 11.165 & 0.96024 & 0.04699\\
    & & & & 60 & 11.877 & 0.96573 & 0.03418\\
    \hline
    0.15 & 2 & 0.001 & 1.82308 & 50 & 9.188 & 0.93928 & 0.01148\\
    & & & & 60 & 9.531 & 0.94344 & 0.00643\\
    & 2 & 0.0001 & 1.82588 & 50 & 9.78 & 0.95294 & 0.02250\\
    & & & & 60 & 10.26 & 0.95795 & 0.01477\\
    \hline
    0 & 4 & 0.001 & 2.85989 & 50 & 20.186 & 0.94692 & 0.22242\\
    & & & & 60 & 22.072 & 0.95624 & 0.17598\\
    & 4 & 0.0001 & 2.86961 & 50 & 20.21 & 0.94201 & 0.30108\\
    & & & & 60 & 22.1 & 0.95163 & 0.24986\\
    \hline
    0.1 & 4 & 0.001 & 2.68355 & 50 & 16.384 & 0.94986 & 0.08656\\
    & & & & 60 & 17.46 & 0.95677 & 0.05945\\
    & 4 & 0.0001 & 2.69192 & 50 & 16.89 & 0.95339 & 0.13961\\
    & & & & 60 & 18.15 & 0.96130 & 0.10619\\
    \hline
    0.15 & 4 & 0.001 & 2.60461 & 50 & 14.774 & 0.94608 & 0.05076\\
    & & & & 60 & 15.56 & 0.95192 & 0.03213\\
    & 4 & 0.0001 & 2.61237 & 50 & 15.38 & 0.95385 & 0.08869\\
    & & & & 60 & 16.36 & 0.96086 & 0.06379\\
    \hline
 \end{tabular}
 \caption{\label{table:2} The e-fold number $N$, the spectral index parameters $n_s$ and $r$ are presented here for the potential $V=V_0(1-\phi^{p})e^{-\lambda\phi}$.}
 \end{table}
While the $n_s - r$ points corresponding to $N=50$ and $60$ are very sensitive to p-value as for the parameter value $p=4$, one can see that most of the data points lie outside the Planck region. For $\xi=0.001$, we see that for the parameter choice, $\lambda =0$, $p=2$ and $\lambda=0.1$, $p=2$, the $n_s - r$ data points corresponding to $N=50$ and $60$ lie well within the Planck data. But as we increase the value of $\lambda $ from $0.1 $ to $0.15$ for $p=2$, the tensor-to-scalar ratio decreases while $n_s$ value shifts beyond $\pm 3\sigma$ range i.e. for this case the $n_s - r$ points are sensitive to $\lambda $ value. Additionally, $p=4$ power law for the hilltop potential may be more intriguing from the perspective of particle physics considering a no-scale renormalizable scalar potential. 

 \noindent
From Fig (\ref{figure2}), it appears that the $p=4$ scenario agrees with the observations quite well for suitable choices of the potential parameters. For the case $\lambda=0.1$, $p=4$ and $\lambda=0.15$, $p=4$, we find that the $n_s$ and the $r$ only matches in $N=60$ case, while for $N=50$, both $n_s$ and $r$ are found to lie beyond the Planck region. In the $p=4$ case, we see that as we increase $\lambda$, the tensor-to-scalar ratio reduces while the $n_s$ value still remains away from the $\pm 3\sigma$ range of the Planck data for the case with the e-fold number $N=50$. \\
Overall, for fixed $\lambda$, by decreasing $\xi$, the $n_s$ value moves more and more towards the central value $0.9649$, which results in an increase in $r$. On the other hand, for a fixed $\xi$, if we increase $\lambda$, we get a smaller $r$ value.

\subsubsection{Case 3: $V = V_0 (1 -\lambda \phi)^p$}
In this section, we will examine the inflationary universe model with the inflation potential $V = V_0 (1-\lambda \phi)^p$, which in  Einstein's frame, takes the following form:
\begin{equation}
    \hat{V}(\phi) = \frac{V_0 \bigl(1 - \lambda \phi\bigr)^p}{\bigl(1 + \xi \phi^2\bigr)^2}
\end{equation}
This specific potential has been considered earlier by Gr{\o}n\cite{Gron} for the minimally coupled gravity to study inflation but it could not produce the necessary e-folds in order to solve the 
flatness and horizon problem which is the main motivation to study this potential in a non-minimal framework.
The slow roll parameters for this potential are calculated as      
\begin{equation}
    \epsilon_v = \frac{\Bigl[-4\xi\phi\bigl(-1+\lambda\phi\bigr)+p\bigl(\lambda+\lambda\xi\phi^2\bigr)\Bigr]^2}{2\bigl(1-\lambda\phi\bigr)^2\bigl\{1+\xi(1+12\xi)\phi^2\bigr\}}
\end{equation}
\begin{equation}
\begin{split}
    \eta_v 
    &=\frac{1}{\bigl(-1+\lambda\phi\bigr)^2\bigl(1+\xi\phi^2+12\xi^2\phi^2\bigr)^2} \times \biggl[p^2\bigl(\lambda+\lambda\xi\phi^2\bigr)^2\bigl(1+\xi\phi^2+12\xi^2\phi^2\bigr)+4\xi\bigl(-1+\\
    &\qquad \lambda\phi\bigr)^2\bigl(-1+3\xi\phi^2+4\xi^2\phi^4+48\xi^3\phi^4\bigr)-p\lambda\bigl(1+\xi\phi^2\bigr)\Bigl\{\lambda\bigl\{1+9\xi\phi^2+96\xi^3\phi^4+8\xi^2  \\
    &\qquad
    \times \phi^2\bigl(3+\phi^2\bigr)\bigr\}-~\xi\phi\bigl\{7+84\xi^2\phi^2+\xi\bigl(12+7\phi^2\bigr)\bigr\}\Bigr\}\biggr]
    \end{split}
\end{equation}
Accordingly, we derive the scalar spectral index  as 
\begin{equation}
\begin{split}
    n_s 
    &= - \frac{1}{\bigl(-1+\lambda\phi\bigr)^2\bigl(1+\xi\phi^2+12\xi^2\phi^2\bigr)^2} \times \biggl[ p^2(\lambda+\lambda\xi\phi^2)^2(1+\xi\phi^2+12\xi^2\phi^2)+\bigl(-1\\
    &\qquad +\lambda\phi\bigr)^2\bigl\{-1-\xi^2\phi^4-8\xi^3\phi^4+48\xi^4\phi^4-2\xi\bigl(-4+\phi^2\bigr)\bigr\}-2p\lambda\bigl(1+\xi\phi^2\bigr)\Bigl\{\lambda\bigl\{-1\\
    &\qquad +3\xi\phi^2+48\xi^3\phi^4+4\xi^2\phi^2\bigl(-6+\phi^2\bigr)\bigr\}-\xi\phi\bigl\{5+60\xi^2\phi^2+\xi\bigl(-12+5\phi^2\bigr)\bigr\}\Bigr\}\biggr]
\end{split}
\end{equation}
and the tensor to scalar ratio and the tensor spectral index  as  
\begin{equation}
    r = \frac{8\Bigl[-4\xi\phi\bigl(-1+\lambda\phi\bigr)+p\bigl(\lambda+\lambda\xi\phi^2\bigr)\Bigr]^2}{\bigl(-1+\lambda\phi\bigr)^2\bigl(1+\xi(1+12\xi)\phi^2\bigr)}, ~~ n_T = -\frac{\Bigl[-4\xi\phi\bigl(-1+\lambda\phi\bigr)+p\bigl(\lambda+\lambda\xi\phi^2\bigr)\Bigr]^2}{\bigl(-1+\lambda\phi\bigr)^2\bigl(1+\xi(1+12\xi)\phi^2\bigr)}
\end{equation}
\begin{table}[H]
\addtolength{\tabcolsep}{5.8pt}
 \centering
 \small
 \begin{tabular}{cccccccccc}
 \hline
     $\lambda$ & p & $\xi$ & $\phi_f$ & $N$& $\phi$ & $n_s$ & $r$\\
     \hline
     1 & 2 & 0.001 & 2.40869 & 50 & 14.52 & 0.95915 & 0.09659\\
     & & & & 60 & 15.67 & 0.96554 & 0.07356\\
     & & 0.0001 & 2.41366 & 50 & 15.15 & 0.96045 & 0.15007\\
     & & & & 60 & 16.46 & 0.96694 & 0.12416\\
     \hline
     -1 & 2 & 0.001 & 0.41268 & 50 & 12.52 & 0.95885 & 0.10115\\
     & & & & 60 & 13.655 & 0.96522 & 0.07748\\
     & & 0.0001 & 0.41406 & 50 & 13.155 & 0.96043 & 0.15080\\
     & & & & 60 & 14.47 & 0.96694 & 0.12476\\
     \hline
     1 & -2 & 0.001 & 2.42811 & 50 & 17.98 & 0.96901 & 0.31273\\
     & & & & 60 & 20.24 & 0.96880 & 0.29270\\
     & & 0.0001 & 2.41559 & 50 & 15.46 & 0.99704 & 0.17064\\
     & & & & 60 & 16.88 & 0.99705 & 0.14455\\
     \hline
     -1 & -2 & 0.001 & 0.41600 & 50 & 15.35 & 0.97181 & 0.29182\\
     & & & & 60 & 17.45 & 0.97135 & 0.27282\\
     & & 0.0001 & 0.41439 & 50 & 13.41 & 0.99733 & 0.16901\\
     & & & & 60 & 14.82 & 0.99731 & 0.14298\\
     \hline
 \end{tabular}
 \caption{\label{table:3} The e-fold number, the spectral index parameters $n_s$ and $r$ are presented here for the potential $V = V_0 (1 -\lambda \phi)^p$.}
 \end{table}
\noindent
In Table~(\ref{table:3}), we have presented $n_s$ and $r$  for fixed value of $\phi$ corresponding to $N=50$ and $60$ as before. We are getting the values of $n_s$ and $r$ consistent with the Planck 2018 data for $\lambda=\pm 1,p=2$ and $\xi=0.001$. However, for all other parameter options, we find the $n_s$ and $r$ values quite larger than the Planck data.

\subsubsection{Case 4: $V=V_0\frac{\alpha\phi^2}{1+\alpha\phi^2}$}
\begin{figure}[H]
 \centerline{\psfig{file=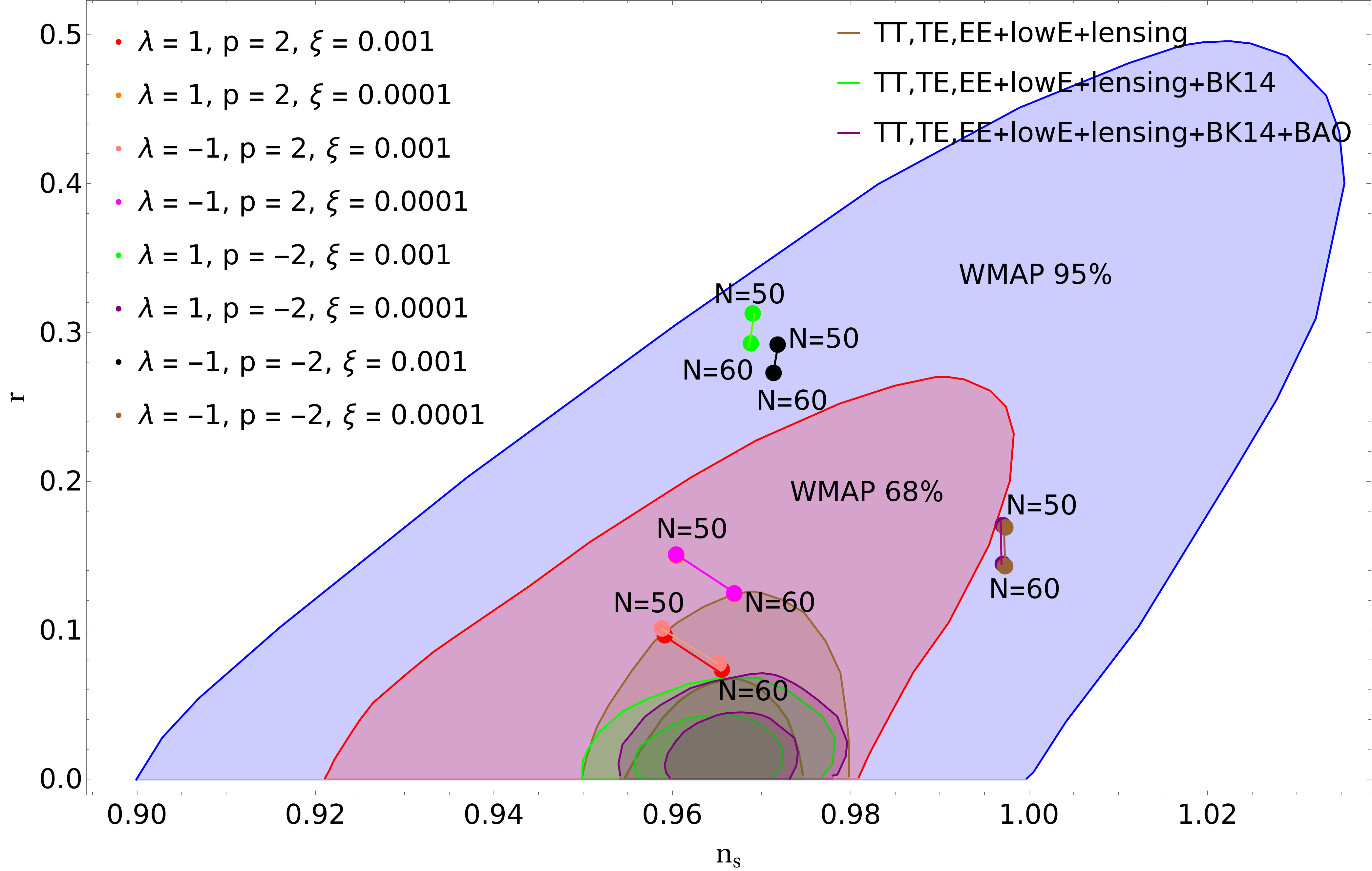, width=8.5cm}\psfig{file=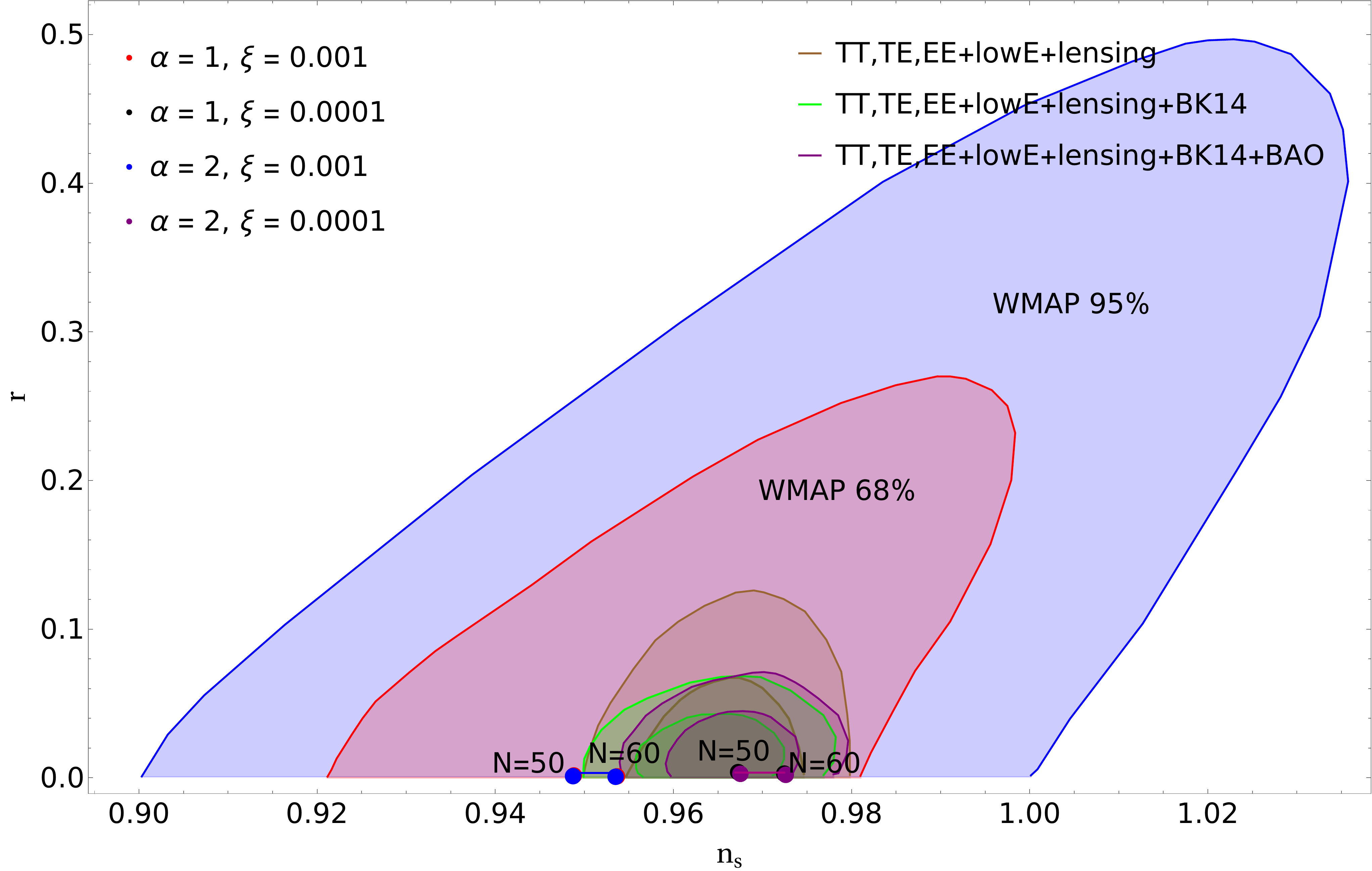, width=8.5cm}}
  \vspace*{5pt}
  \caption{\label{figure3}{(Color online) Constraints on $n_s$ and $r$ from CMB measurements for two different potentials $V = V_0 (1 -\lambda \phi)^p$ and $V=V_0\frac{\alpha\phi^2}{1+\alpha\phi^2}$ for different $\lambda$, p and $\xi$ values. Shaded regions are allowed by WMAP measurements, PLANCK alone, PLANCK+BK15, PLANCK+BK15+BAO to $68\%$ and $95\%$ confidence level. 
  }}
\end{figure}

The last potential in our consideration is $V=V_0\frac{\alpha\phi^2}{1+\alpha\phi^2}$. In Einstein's frame, it takes the form 
\begin{equation}
\hat{V}(\phi)=\frac{V_0\alpha\phi^2}{\bigl(1+\alpha\phi^2\bigr)\bigl(1+\xi\phi^2\bigr)^2}
\end{equation}
where $V_0$ is a constant, $\lambda$ is the potential parameter. This fractional potential was first studied by Eshagli et al.\cite{Eshagli}.
The two slow-roll parameters, in this case, are found to be,
\begin{equation}
    \epsilon_v=\phantom{-}\frac{2\bigl[-1+\xi\bigl(\phi^2+2\alpha\phi^4\bigr)\bigr]^2}{\bigl(1+\xi\phi^2+12\xi^2\phi^2\bigr)\bigl(\phi+\alpha\phi^3\bigr)^2}
\end{equation}
\begin{equation}
\begin{split}
    \eta_v
    & =\frac{1}{\phi^2\bigl(1+\alpha\phi^2\bigr)^2\bigl(1+\xi\phi^2+12\xi^2\phi^2\bigr)^2} \biggl[2\Bigl\{1-6\xi\phi^2-5\xi^2\phi^4+24\xi^4\phi^6+2\xi^3\phi^4\bigl(-36 \\
   &\quad +\phi^2\bigr)+2\alpha^2\xi\phi^6\bigl(-1+3\xi\phi^2+4\xi^2\phi^4+48\xi^3\phi^4\bigr)+\alpha\phi^2\bigl\{-3-20\xi\phi^2+72\xi^4\phi^6+\\
   &\quad
   6\xi^3\phi^4\bigl(-28+\phi^2\bigr)-\xi^2\phi^2\bigl(48+11\phi^2\bigr)\bigr\}\Bigr\}\biggr]
\end{split}
\end{equation}
\vspace*{-0.1in}
Accordingly, we find the scalar spectral index as,
\begin{equation}
    \begin{split}
        n_s
        &= \frac{1}{\phi^2\bigl(1+\alpha\phi^2\bigr)^2\bigl(1+\xi\phi^2+12\xi^2\phi^2\bigr)^2} \times \biggl[\Bigl\{-8-\bigl(-1+12\alpha+12\xi+144\xi^2\bigr)\phi^2+\\
        &\qquad 2\phi^4\bigl(\alpha+\xi-16\alpha\xi+8\xi^2-96\alpha\xi^2\bigr)+\alpha^2\bigl(1-8\xi\bigr)+4\xi\alpha\bigl(1+\xi-24\xi^2\bigr)+\xi^2\bigl(1\\
        &\qquad +20\xi+96\xi^2\bigr)\Bigr\}\phi^6+2\alpha\xi\phi^8\bigl(\alpha+\xi+12\xi^2\bigr)+\alpha^2\xi^2\phi^{10}\bigl(1+8\xi-48\xi^2\bigr)\biggr]
    \end{split}
\end{equation}

\noindent 
and the tensor-to-scalar ratio and the tensor spectral index as 

\begin{equation}
    r = \frac{32\bigl[-1+\xi\bigl(\phi^2+2\alpha\phi^4\bigr)\bigr]^2}{\bigl(1+\xi\phi^2+12\xi^2\phi^2\bigr)\bigl(\phi+\alpha\phi^3\bigr)^2},
    ~~ n_T = -\frac{4\bigl[-1+\xi\bigl(\phi^2+2\alpha\phi^4\bigr)\bigr]^2}{\bigl(1+\xi\phi^2+12\xi^2\phi^2\bigr)\bigl(\phi+\alpha\phi^3\bigr)^2}
\end{equation}
\noindent 
We have estimated $n_s$ and $r$ for fixed values of $\xi (=0.001,0.0001)$ corresponding to the field value $\phi$ which is obtained by taking $N=50$ and $60$ as shown in Table~\ref{table:4}. For $\alpha=1$ and $2$, we see that both $n_s$ (up to $\pm 3\sigma$ C.L.) and $r$ agree well with the PLANCK 2018 data. Although, there is an exception for $\xi=0.001$ corresponding to $N=50$.
\begin{table}[H]
\addtolength{\tabcolsep}{9.9pt}
 \centering
 \small
 \begin{tabular}{ccccccccc}
 \hline
     $\alpha$ & $\xi$ & $\phi_f$ & $N$& $\phi$ & $n_s$ & $r$\\
     \hline
     1& 0.001 & 0.83312 & 50 & 3.993 & 0.948824 & 0.001556\\
     & & & 60 & 4.118 & 0.953659 & 0.00096\\
     & 0.0001 & 0.83395 & 50 & 4.326 & 0.967264 & 0.003783\\
     & & & 60 & 4.528 &  0.97244 & 0.002813\\
     \hline
     2 & 0.001 & 0.70649 & 50 & 3.378 & 0.948731 & 0.00107\\
     & & & 60 & 3.481 & 0.953508 & 0.00065\\
     & 0.0001 & 0.70705 & 50 & 3.664 & 0.967517 & 0.002635\\
     & & & 60 & 3.832 & 0.972611 & 0.001964\\
     \hline
 \end{tabular}
 \caption{\label{table:4} The e-fold number, the spectral index parameters $n_s$ and $r$ are presented here for the potential $V=V_0\frac{\alpha\phi^2}{1+\alpha\phi^2}$.}
 \end{table}
\noindent  
For fixed $\alpha$, if we decrease $\xi$ we will get higher $n_s$ and $r$ values. On the other hand for fixed $\xi$, increasing the $\alpha$ value decreases $r$ slightly although $n_s$ remains unchanged upto third decimal place.\\
\noindent Similarly in Fig (\ref{figure3}), we have shown $n_s-r$ plot for potential $V=V_0(1-\lambda\phi)^p$ (left side) and for $V=V_0\frac{\alpha\phi^2}{1+\alpha\phi^2}$ (right side). From the left plot, we can observe that for $p=-2$ the calculated $n_s$ and $r$ lie in $95\%$ WMAP data range but for rest of the values, they are lying within WMAP $68\%$ even in the PLANCK region. On the other hand, $V=V_0\frac{\alpha\phi^2}{1+\alpha\phi^2}$ produces $n_s,r$ values which are matching mostly with PLANCK data as observed from the right panel of Fig (\ref{figure3}) except for the $N=50$ data point for $\alpha=1$, $2$ and $\xi=0.001$.
\subsection{Case 5: (i) $V=V_0 \phi^p e^{-\lambda\phi}$, (ii) $V=V_0 (1- \phi^p) e^{-\lambda\phi}$, (iii) $V = V_0 (1 -\lambda \phi)^p$ and (iv) 
 $V=V_0\frac{\alpha\phi^2}{1+\alpha\phi^2}$ with $\xi=0$}

\noindent
In the table (\ref{table:5}), we have presented the different cosmological parameter values for all four potentials in minimal coupling case $(\xi=0)$. We can notice that for $V\propto \phi^p e^{-\lambda\phi}$, the obtained $n_s$ value is outside of the PLANCK and WMAP data region, and the $r$ value, in this case, is likewise very high.  \\
After comparing the tables (\ref{table:1}) and (\ref{table:5}), 
we can infer that after the inserting non-minimal parameter $\xi$, the $n_s$ value fits the $\pm 3 \sigma $ C.L. of PLANCK 2018 and $r<0.106$.
Similar to this, for $V\propto (1-\lambda\phi)^p$, $n_s$ value matches with the central value of observational data and $r>0.106$ for the minimal case whereas by taking $\xi\neq 0$ makes $r<0.106$ as shown in the table (\ref{table:2}). On the other hand, for $V\propto (1-\phi^p)e^{-\lambda\phi}$ and $V\propto \frac{\alpha\phi^2}{1+\alpha\phi^2}$, $n_s$ value lies within $\pm 1\sigma$ C.L. of PLANCK data along with $r<0.106$ but for $\xi\neq 0$ the calculated $n_s$ value lies within $\pm 3\sigma$ C.L., also $r<0.106$ lies in the PLANCK 2018 data range. 
 
\begin{table}[H]
\addtolength{\tabcolsep}{8.6pt}
  \small
 \begin{tabular}{ccccccccc}
 \hline
 Potential &$p$ &$\lambda$ & $\phi_f$ & $N$& $\phi$ & $n_s$ & $r$ \\
    \hline
    $V_0\phi^p e^{-\lambda\phi}$ & 2 & 0.1 & 1.32082 & 50 & 5.456 & 0.794568 & 0.568472\\
    & & & & 60 & 6.163 & 0.844281 & 0.403264\\
    \hline
    $V_0(1-\phi^p)e^{-\lambda\phi}$ & 2 & 0.1 & 1.85880 & 50 & 11.243 & 0.96131 & 0.05032\\
    & & & & 60 & 11.98 & 0.96690 & 0.03712\\
    \hline
    $ V_0(1-\lambda\phi)^p$& 2 & 1 & 2.41421 & 50 & 15.21 & 0.96038 & 0.15848\\
    & & & & 60 & 16.56 & 0.96696 & 0.13217\\
    \hline
    Potential & $\alpha$ &  & $\phi_f$& $N$ &  $\phi$ &$n_s$ & $r$  \\
    \hline
    $V_0\frac{\alpha\phi^2}{1+\alpha\phi^2}$ & 1 &  & 0.834040 & 50 & 4.368 & 0.969197 & 0.004160 \\
    & & & & 60 & 4.58 & 0.974364 & 0.003158\\
    \hline
 \end{tabular}
 \caption{\label{table:5} The e-fold number $N$ and the spectral index parameters $n_s$, $r$ and $n_T$ are presented here for all four potentials in minimal scenario $(\xi=0)$.}
 \end{table}  
 \noindent
As a result, we may say that $\xi\neq 0$ broadens the $n_s$ data range and produces $r<0.106$, which supports our decision to analyze the aforementioned potentials in a non-minimal framework.

\subsection{Cosmological Viability}
\noindent
In this manuscript, we have studied inflation in the non-minimal coupled gravity where we have presumed the coupling function between the scalar field and the gravity is $\frac{1}{2}\xi R\phi^2$. We have considered different inflaton potentials to study non-minimal inflation. We first present the generic formulations for the slow-roll parameters as well as the spectral index parameters and then examine the impact of the coupling term $\frac{1}{2}\xi R\phi^2$ on the potentials.
To check whether the model is cosmologically viable, we have compared it with the nine-year WMAP and the most recent PLANCK 2018 data. This model with non-minimal coupling is consistent with Planck 2018 TT, TE, and EE $+$ lowE $+$ lensing which is further tightened by the BK14 and BAO data, for some particular choices of parameters, which are discussed in detail in the section of different potentials. Our analysis demonstrates that in the non-minimal coupling model, with the small values of the parameter, $\xi$ is consistent with the observational data. Though, it is viable only in the weak coupling limit ($\sim 10^{-3}, 10^{-4}$).

\noindent {\it {\bf Comparison of our work with previous studies:}}
In Ref. \cite{Nozari}, Nozari {\it et al} have considered quadratic potential, as well as exponential potential in the non-minimal coupling theory and their estimate of $n_s$ and $r$, agrees well with the WMAP3 data. The range of $\xi$ for two potentials are shown in Table \ref{table:6}. Similarly, in \cite{Nozari2}, DGP Brane-like inflationary scenario has been studied in non-minimal framework and the authors found the constraint on non-minimal coupling parameter from WMAP3 data as $\xi\leq -1.05\times 10^{-1}$ and $\xi\geq 1.2\times10^{-2}$.
In \cite{Hrycyna}, the authors have studied non-minimal coupling in the context of Dynamical system analysis in the Jordan frame where they have taken a system of autonomous equations to show the current accelerated expansion of the universe. They have done the analysis for both canonical scalar field and Phantom scalar field and obtained the constraints on $\xi$ for both cases which are shown in Table \ref{table:6}.


\begin{table}[H]
\addtolength{\tabcolsep}{-2pt}
 \small
 \begin{tabular}{c|c|c|c}
 \hline
 Sr. No. & Model & Authors & Range on $\xi$ \\
 \hline
 1. & $V(\phi) = \lambda \phi^4$ & K. Nozari et al\cite{Nozari} & $\xi \le -0.1666, \xi \ge 0.01$\\

 2. & $V(\phi) = V_0~exp\Bigl( -\sqrt{\frac{16 \pi}{p m_{pl}^2}\phi} \Bigr)$ & K. Nozari et al\cite{Nozari} & $0.271 \le \xi \le 0.791$\\

 3. & Canonical scalar field & O. Hrycyna et al\cite{Hrycyna} & $\xi=0.1837_{-0.0598}^{+0.0513}$ \\

 4. & Phantom scalar field & O. Hrycyna et al\cite{Hrycyna} & $\xi=0.2449_{-0.0624}^{+0.0694}$ \\

 5. & DGP Brane Inflation & K. Nozari et al\cite{Nozari2} & $\xi \ge 1.2 \times 10^{-2}$ \\

 6. & Higgs like Inflation & T. Takahashi and T. Tenkanen\cite{Takahashi,Tenkanen} & $\xi= 0.01, 0.03, 0.05$ \\
 7. & Quadratic model & L. Boubekeur {\it et al.}\cite{Boubekeur} & $0.0028_{-0.0025}^{+0.0023}<\xi<0.0024_{-0.0023}^{+0.0023}$\\
 8. & $\xi_2\phi^2+\xi_4\phi^4$ & Glavan {\it et al.}\cite{Glavan} & $\xi_4 \in (-0.1,-.01), \xi_2\in(-0.005,-0.002)$\\
 9. & Our model & P. Sarkar et al & $10^{-4}<\xi<10^{-3}$\\
 \hline
 \end{tabular}
 \caption{\label{table:6} comparison with previous non-minimal models.}
 \end{table}  
 \noindent
In \cite{Boubekeur}, the author has taken a non-minimally coupled inflationary model in the Einstein frame and matched the cosmological parameters data with Planck 2015 results for quadratic potential. They have emphasized on both positive and negative values of $\xi$ and shown that for a particular choice of $\xi$, $n_s \approx 0.96$. In \cite{Glavan}, the authors have regarded the non-minimal coupling between a scalar field and gravity as $(\xi_2 \phi^2 + \xi_4 \phi^4) R$ in the Einstein frame. Their model is consistent with the observational data only for $N=65$ with a few maximal values of $n_s$, while is detrimental for the $N=50$ and $60$ case. In addition to allowing the action to have an $R^2$ term in the context of Palatini gravity, the authors have explored a Higgs inflation model in which the Standard Model Higgs couples non-minimally to gravity \cite{Takahashi, Tenkanen}. They have taken different values of $\xi=0.01,0.03,0.05$ and shown that in the Palatini version of Higgs inflation, the tensor-to-scalar ratio is of the order of $\sim 10^{-5}$. \\
An interesting aspect of our model is that it can possibly explain the non-Gaussianity present in the CMBR data - to understand its origin and how to deal with it with the proper statistics against the Gaussian Noise background of the CMBR data is a challenging task and a matter of intense research activity at present. It may arise in the CMBR data if the primordial fluctuations themselves are non-Gaussian in nature. The non-Gaussian signatures are expected in a broad range of inflation models that have multiple fields or non-minimal derivative couplings. The inflation model with a non-minimal coupling $\frac{1}{2}\xi R \phi^2$ which we have considered is likely to produce a certain amount of non-Gaussianity and it is possible to make an estimate of $f_{NL}$, which quantifies the non-Gaussianity\footnote{The simplest single-field, slow-roll inflation models predict nearly Gaussian initial density fields of perturbation. Deviations from Gaussianity are usually parameterized by a non-Gaussian potential given by $$\Psi_{NG}(x) =  -\Bigl(\phi_g(x) + f_{NL}[\phi^2_g(x) - \langle \phi^2_g \rangle] \Bigr)$$ 
where $f_{NL}=$ constant that measures the nature and size of non-Gaussianity. Here $\Psi_{NG}$ is the Newtonian potential and $\phi_g$ is the Gaussian random field (e.g., Salopek $\&$ Bond 1990\cite{Salopek}, Matarrese, Verde and Jimenez 2000\cite{Matarrese}, Komatsu and Spergel 2001\cite{Komatsu1}).
 A measure of non-Gaussianity $ f_{NL} = \sim 10^{-2}$ (For simple inflaton model (Maldacena 2003\cite{Maldacena})), while $f_{NL} = 30 \pm 20$ at the $68\%$ confidence level (e.g., Komatsu 2010\cite{Komatsu2})  }
 as a function of the non-minimal parameter $\xi$. However, a detailed analysis of this is beyond the scope of the present work and will be reported in future work by the authors.

\section{Conclusion:}
In this work, we have studied the inflationary expansion using a class of  inflaton potentials (i) $V\propto\phi^pe^{-\lambda\phi}$, (ii)$V\propto (1-\phi^p)e^{-\lambda\phi}$,(iii)$V\propto (1-\lambda\phi)^p$ and (iv)$V\propto \frac{\alpha\phi^2}{1+\alpha\phi^2}$ within the framework of non-minimal gravity theory. We introduce a non-minimal curvature-scalar mixing term of the form $\frac{1}{2}\xi R \phi^2$ and found that the model can predict $r < 0.106$ (PLANCK+BAO) and $n_s$ within $\pm 3 \sigma$ of the Planck 2018 data and WMAP measurements for a wide range of potential parameter space corresponding to $\xi=0.001$  and $\xi=0.0001$.
We found that the potential $V=V_0\frac{\alpha\phi^2}{1+\alpha\phi^2}$ fits best in our non-minimal gravity framework to produce inflation. We have compared our results with other previous studies and found that the non-minimal parameter $\xi=10^{-3}$ and $10^{-4}$ are comparable with some of the existing studies.
\section*{Acknowledgment}
PS would like to thank the Department of Science and Technology, Government of India for INSPIRE fellowship. Ashmita would like to thank the BITS Pilani K K Birla Goa campus for the fellowship support.

\end{document}